# Quantum melting and Pomeranchuk effect of zero-field Wigner solids

**Boliang Liu[1], Simon Munyan[1], Sina Ahadi[1], Erez Berg[1], and Susanne Stemmer [1,*]**

[1]Materials Department, University of California, Santa Barbara, CA 93106-5050, USA

[*] Corresponding author: stemmer@mrl.ucsb.edu

**ABSTRACT**

Wigner solids and crystals are many-body phases of matter in which repulsive interactions between the charge carriers cause them to freeze into a solid. How Wigner solids compete or cooperate with integer and fractional quantum Hall liquids has been a long-standing open question. Here, we investigate this question for zero-field, electron and hole Wigner solids in cadmium arsenide quantum wells. In the low temperature limit, our results are consistent with a recent theoretical model [A. P. Reddy and L. Fu, Phys. Rev. B **113**, L161403 (2026)] that proposes that the quantum Hall liquid can become energetically favored over the solid at low integer filling factors. Going beyond the low-temperature limit, we observe a re-entrant solid that emerges from the quantum Hall liquid. We explain this observation by considering the role of spin entropy, in close analogy with the zero-field Pomeranchuk effect, but appearing here at finite fields. The competition between quantum Hall liquids and Wigner solids is thus governed by both energy and entropy terms, giving rise to rich phase stability diagrams that result from the interplay between topology and correlations.

## Introduction

Coulomb interactions can cause the electrons in a two-dimensional electron system to crystallize into a lattice, forming a Wigner crystal [1]. Wigner crystals and solids are usually realized in high magnetic fields [2-5], which quench the kinetic energy. Wigner solids have, however, also been reported to exist even in the absence of a magnetic field [6-15]. In some cases, the magnetic field even appears to *destabilize* the Wigner solid [9-11], possibly because of a competition with emerging quantum Hall liquids. Indeed, numerous experiments have demonstrated transitions between Wigner solids and integer/fractional quantum Hall liquid as a function of carrier density, magnetic and electric fields [2-5,9,16-19]. Coexistence of Wigner solids and quantum (anomalous) Hall liquids has also been suggested, including a reentrant quantum Hall effect and anomalous Hall crystals [20-22]. Revealing how Wigner solids compete with topological states of matter in magnetic fields would impact our understanding of the interplay between topology and strong correlations.

Here, we report on the competition between quantum Hall liquids and recently discovered [11] zero-field Wigner solids in cadmium arsenide ($Cd_3As_2$) quantum wells. At small quantum well thicknesses (< 18 nm), electron and hole Wigner solids emerge as gapped states near the bottom of conduction and valence subbands, respectively, where they flank the single-particle, charge neutrality gap. The Wigner solids are not present in thicker quantum wells [23]. DC current-voltage (*I*-*V*) traces exhibit the characteristic signatures of depinning of the Wigner solid's domains from the disorder potential. These characteristics include an abrupt and hysteretic onset of ohmic conduction above a critical voltage bias, which signifies the onset of sliding of the domains.

One of the most remarkable features of these Wigner solids is that they abruptly transition into emerging integer quantum Hall liquids in magnetic fields. Here, we explore this transition over a wide parameter space of magnetic fields, carrier densities, and temperature. In the low temperature limit, our results can be elegantly explained by an energy competition between the solid and the quantum Hall liquid, as recently proposed by Reddy and Fu [24]. Moving beyond the zero-temperature physics, we observe the reentrance of the electron solid at elevated temperatures and intermediate magnetic field strengths. We propose that this transition is governed by spin entropy, in close analogy to the zero-field Pomeranchuk effect.

## RESULTS

We investigate a $Cd_3As_2$ quantum well heterostructure that is similar to those studied in ref. [11], where zero-field electron and hole Wigner solids were first reported. A Landau level map, recorded at a bath temperature of 20 mK, is shown in Fig. 1(a). The map shows the longitudinal conductivity ($\sigma_{xx}$) as a function of gate voltage ($V_g$), which controls the carrier density in the quantum well, and magnetic field ($B$). The minima in $\sigma_{xx}$, i.e., the quantum Hall plateaus, are labeled by their filling factors $\nu$ (see also Supplementary Figure 1). The single-particle charge neutrality gap (indicated by the blue circle) corresponds to $\nu = 0$. The thickness of this quantum well is very near the critical thickness for closing of the topological gap, as can be seen from extrapolating the two zeroth Landau levels to $B = 0$. The electron and hole Wigner solids (green and orange circles, respectively) are bordering the charge neutrality gap. At this thickness, the three high-resistivity regions (Wigner solids and charge neutrality gap) are nearly merged into a single region of low $\sigma_{xx}$ (dark red in the map). For other thicknesses, see ref. [11]. As described in detail in ref. [11], hole and electron Wigner solids can be distinguished from other high-resistivity regions by their distinct $I$-$V$ traces that reflect the pinning/depinning of the solids' domains. Because the $I$-$V$ traces were extensively discussed in ref. [11], we only show a few representative $I$-$V$ traces that serve to distinguish the Wigner solids from the charge neutrality gap in Fig. 1(b). The DC *I-V* traces are recorded at a bath temperature of 12 mK, with $V_g$ tuned into the electron and hole Wigner solids, respectively. Hysteretic and sharp threshold voltage behaviors for ohmic conduction are observed. We refer to ref. [11] for a detailed discussion of these features. A complete set of *I-V* data over a wider range of $V_g$ is shown in the Supplementary Figure 2.

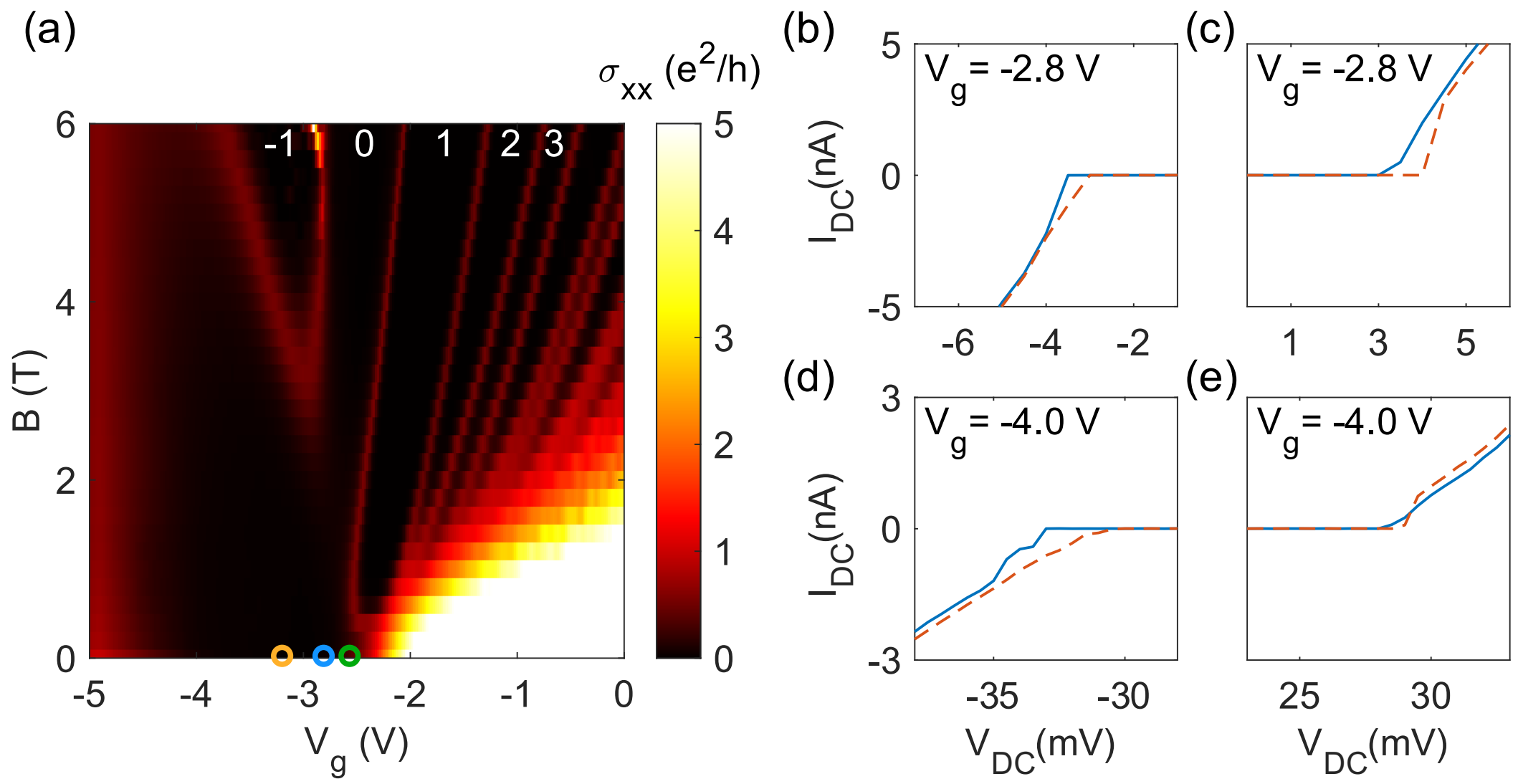

**Figure 1. Integer quantum Hall states and Wigner solids.** (a) Landau level map recorded at a bath temperature of 20 mK in magnetic fields (*B*) up to 6 T. Labels indicate the filling factors (ν) in the Hall plateaus, obtained from the transverse resistivity values. The circles indicate the approximate gate voltage ($V_g$) values at the centers of the charge neutrality gap (blue) and the electron (green) and hole (orange) Wigner solids, respectively. (b-e) DC *I-V* traces showing typical characteristics of the Wigner solids, including threshold voltage and hysteretic behavior, which indicate the onset of sliding of the solids' domains. The *I-V* traces were recorded at a bath temperature of 12 mK at $V_g$ = -2.8 V (electron Wigner solid) and $V_g$ = -4 V (hole Wigner solid), respectively.

Returning to Fig. 1(a), a remarkable feature of the Wigner solids is their abrupt transition into (emerging) quantum Hall liquids. At low fields, the Wigner solids consume the zeroth Landau levels. The zeroth Landau levels emerge at higher fields, along with the $\nu = \pm 1$ Hall liquids, marking the stability range of the Wigner solids in magnetic fields. The transition to the emerging quantum Hall liquid is especially clear for the hole Wigner solid, which survives to much larger fields than the electron Wigner solid. The transition of the electron Wigner solid into the quantum Hall liquid can be clearly seen in the data shown in Fig. 2, discussed below. Understanding the transition from the Wigner solids into quantum Hall liquids in magnetic fields will be a main focus of this paper.

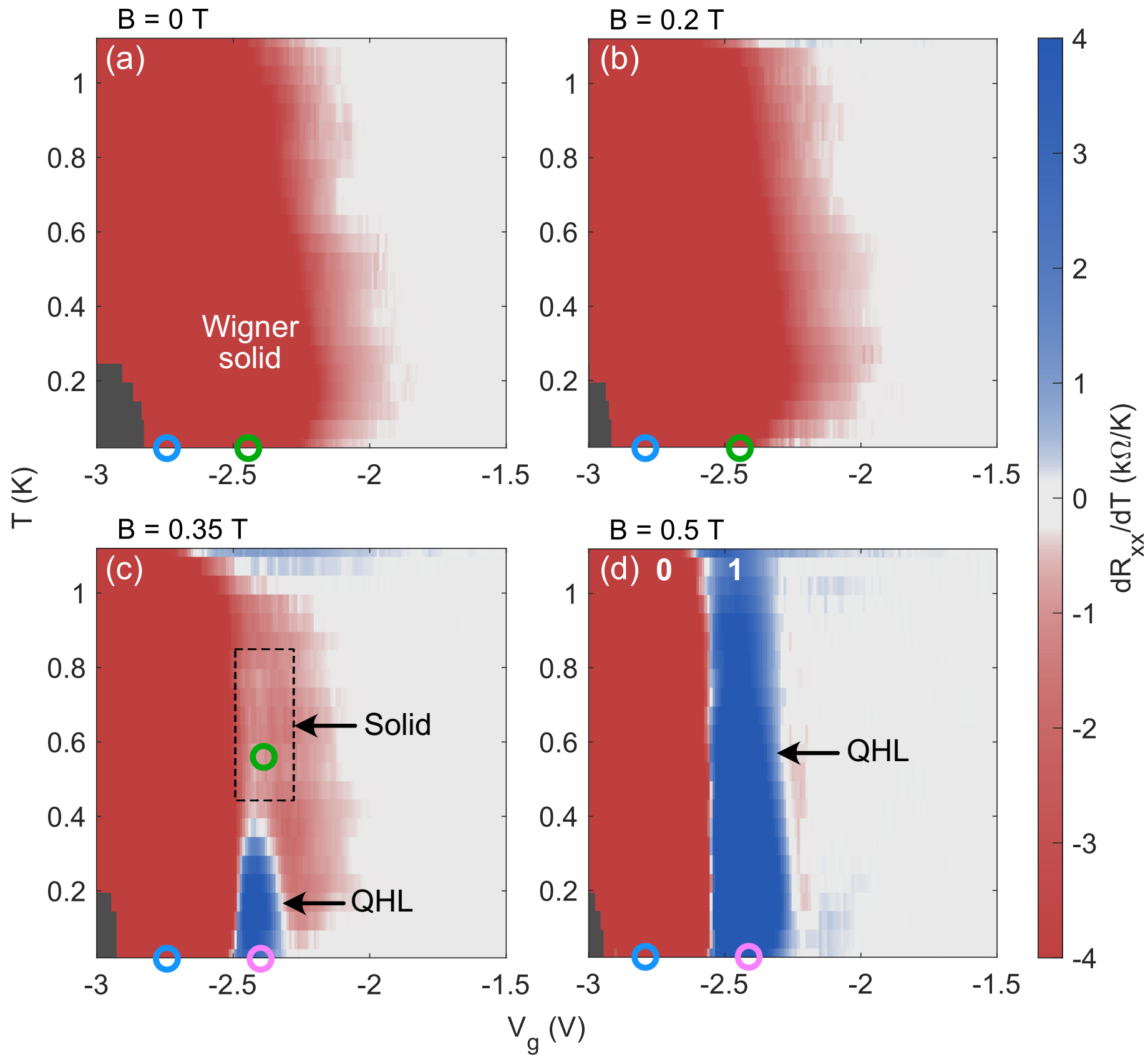

**Figure 2. Competition between the electron Wigner solid and the $\nu = 1$ quantum Hall liquid (QHL).** Shown are maps of the temperature-derivative of the longitudinal AC resistivity ($R_{xx}/dT$) as a function of temperature at gate voltages ($V_g$) that span the charge neutrality gap and electron Wigner solid at different magnetic fields ($B$). (a) $B$ = 0, (b) $B$ = 0.2 T (c) $B$ = 0.35 T, and (d) $B$ = 0.5 T. Circles indicate the charge neutrality gap (blue), electron Wigner solid (green) and emerging Hall liquid (pink), which appears near $\nu = 1$ in (C) and (D), where it replaces the electron Wigner solid. In (C) the solid can be seen to reenter at an elevated temperature (see green circle and dashed rectangle). QHL = quantum Hall liquid. Labels in (d) indicate the filling factors ($\nu$), obtained from the transverse resistivity values.

To distinguish between electron/hole solids and quantum Hall liquids, we turn to the temperature-derivative of the AC longitudinal resistivity ($dR_{xx}/dT$) as a function of carrier density ($V_g$), magnetic field, and temperature ($T$). While transport in sliding Wigner solids and their melting behavior is still a subject of on-going research, they generally exhibit a negative temperature coefficient, i.e., $dR_{xx}/dT < 0$ [16,20,21,25]. Here, we refer to both the pinned and sliding Wigner solids as "solids" to distinguish them from the Hall liquids. In the Supplementary Information, we show that in the Wigner solids the resistance extracted from DC *I-V* data depends exponentially on $T$, consistent with thermal activation. In contrast, $dR_{xx}/dT$ of the emerging Hall liquids follows the known behavior of $R_{xx}$ in the quantum Hall regime, $R_{xx} \propto \exp(-\Delta/k_B T)$, which applies as long as the temperature is not too low [26]. Thus $dR_{xx}/dT > 0$ in the quantum Hall liquids. The different signs of $dR_{xx}/dT$ serve to distinguish the emerging quantum Hall liquids and (sliding) Wigner solids in the temperature range shown in Figs. 2 - 4. This also allows us to determine the phase stability over a wider temperature range, because the sharp threshold behavior seen in Fig. 1(b) only exists as long as the Wigner solids are pinned (also see ref. [11]). We note further that, unlike in the DC measurement in Fig. 1(b), in a current-biased AC measurement the Wigner solids are always forced into the sliding regime (also see Methods section). The raw AC resistivity data used to calculate $dR_{xx}/dT$ are shown in the Supplementary Figures 3 and 4. While nonlinear *I-V* characteristics can make accurate AC voltage measurements difficult, the temperature dependence of the resistivity is determined correctly, as can be seen by comparisons with the DC data (see Supplementary Figures 5 and 6). At elevated temperatures, which is also the temperature regime of interest for the reentrant Wigner solid discussed below, the *I-V* relationship is linear (see Supplementary Figure 6).

Figure 2 shows $dR_{xx}/dT$ as a function of temperature as $V_g$ is tuned from the conduction subband into the charge neutrality gap. No data are presented in the dark grey regions for which the resistivity was too high to accurately measure (see Supplementary Information). The charge neutrality gap and the electron Wigner solid are labeled by blue and green circles, respectively. The red colored regimes indicate $dR_{xx}/dT < 0$, signifying the electron Wigner solid and charge neutrality gap, respectively. In the conduction band, $dR_{xx}/dT \sim 0$ (white region). At zero field (Fig. 2(a)) the electron solid persists to at least 1.2 K. Figures 2(b-d) show the evolution of the electron Wigner solid into the quantum Hall liquid with increasing magnetic field. The precursor to the $\nu = 1$ quantum Hall liquid is the blue stripe ($dR_{xx}/dT > 0$) that first appears at 0.35 T in the low temperature regime (Fig. 2(c)). Importantly, the new liquid phase consumes the electron solid only near $\nu = 1$ ($V_g \approx -2.5$ V), see green (solid) and pink (liquid) circles. The $\nu = 1$ quantum Hall plateau is fully developed at higher fields (see Fig. 3).

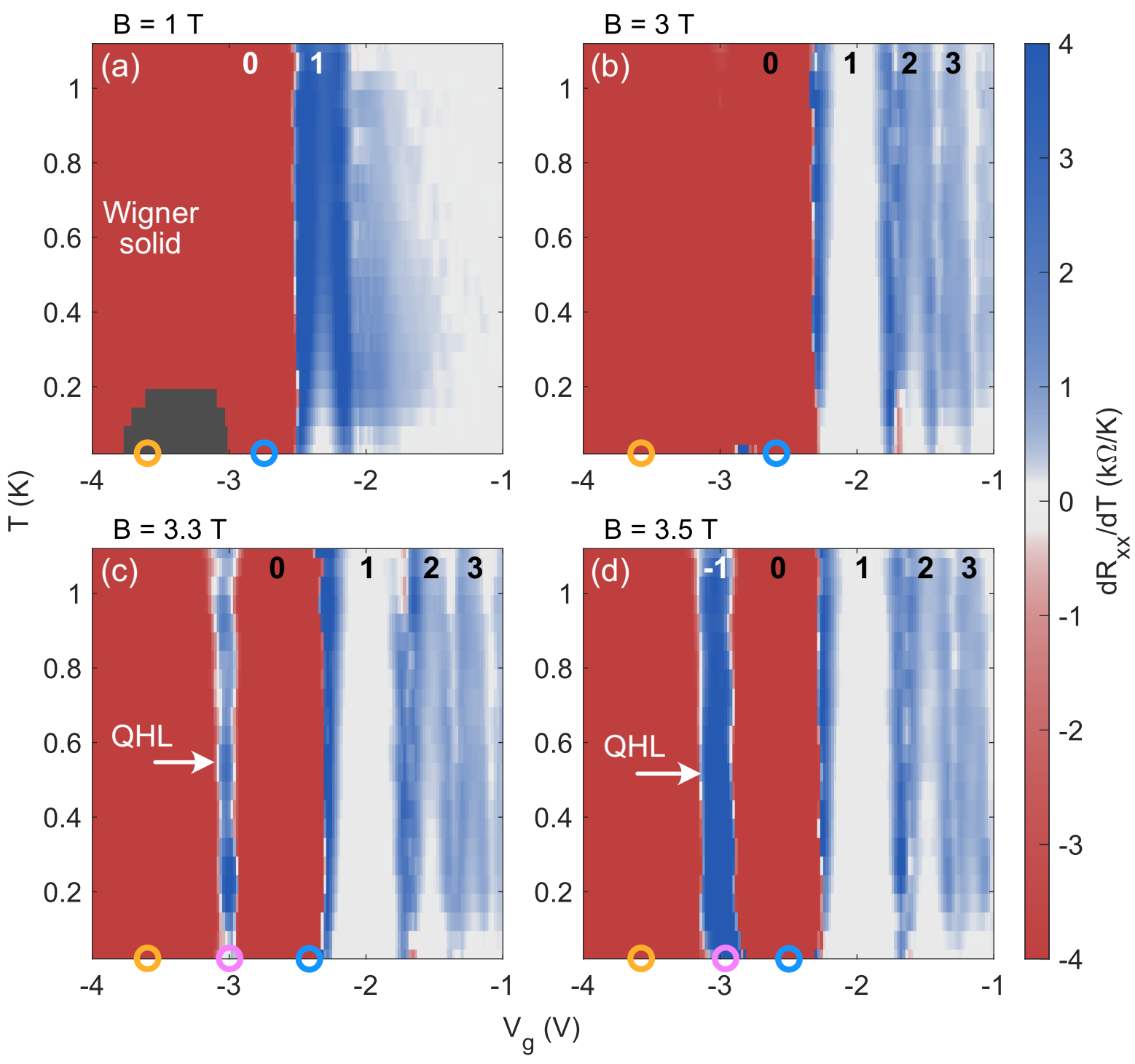

**Figure 3. Competition between the hole Wigner solid and $\nu = -1$ quantum Hall liquid (QHL).** Shown are maps of the temperature-derivative of the longitudinal resistivity ($R_{xx}/dT$) as a function of the temperature and gate voltage ($V_g$) at different magnetic fields ($B$). (a) $B$ = 1 T, (b) $B$ = 3 T (b) $B$ = 3.3 T, and (d) $B$ = 3.5 T. Circles indicate the charge neutrality gap (blue), hole Wigner solid (orange) and emerging Hall liquid (pink), which appears near $\nu = -1$ in (c) and (d), as also indicated by the arrows. Note that the fields are higher than in Fig. 2. Hall plateaus are already fully developed on the electron side at these fields. Labels in the top of each panel indicate the filling factors (ν), obtained from the transverse resistivity values.

Figure 3 shows higher field data that includes the valence subband. The $\nu = -1$ quantum Hall plateau emerges from the hole Wigner solid near $V_g = -3$ V for $B > 3$ T. At high fields, when both Wigner solids have been consumed, the $\nu = 1$ and $\nu = -1$ quantum Hall plateaus define the boundaries of the charge neutrality gap. Because the Landau levels disperse in field, the charge neutrality gap widens and shifts to more towards positive $V_g$ as $B$ increases (see also Fig. 1(a)). Even as $\nu = -1$ becomes fully quantized (see transverse conductivity data presented in the Supplementary Information), however, the hole solid still persists at more negative $V_g$ ($\nu << -1$).

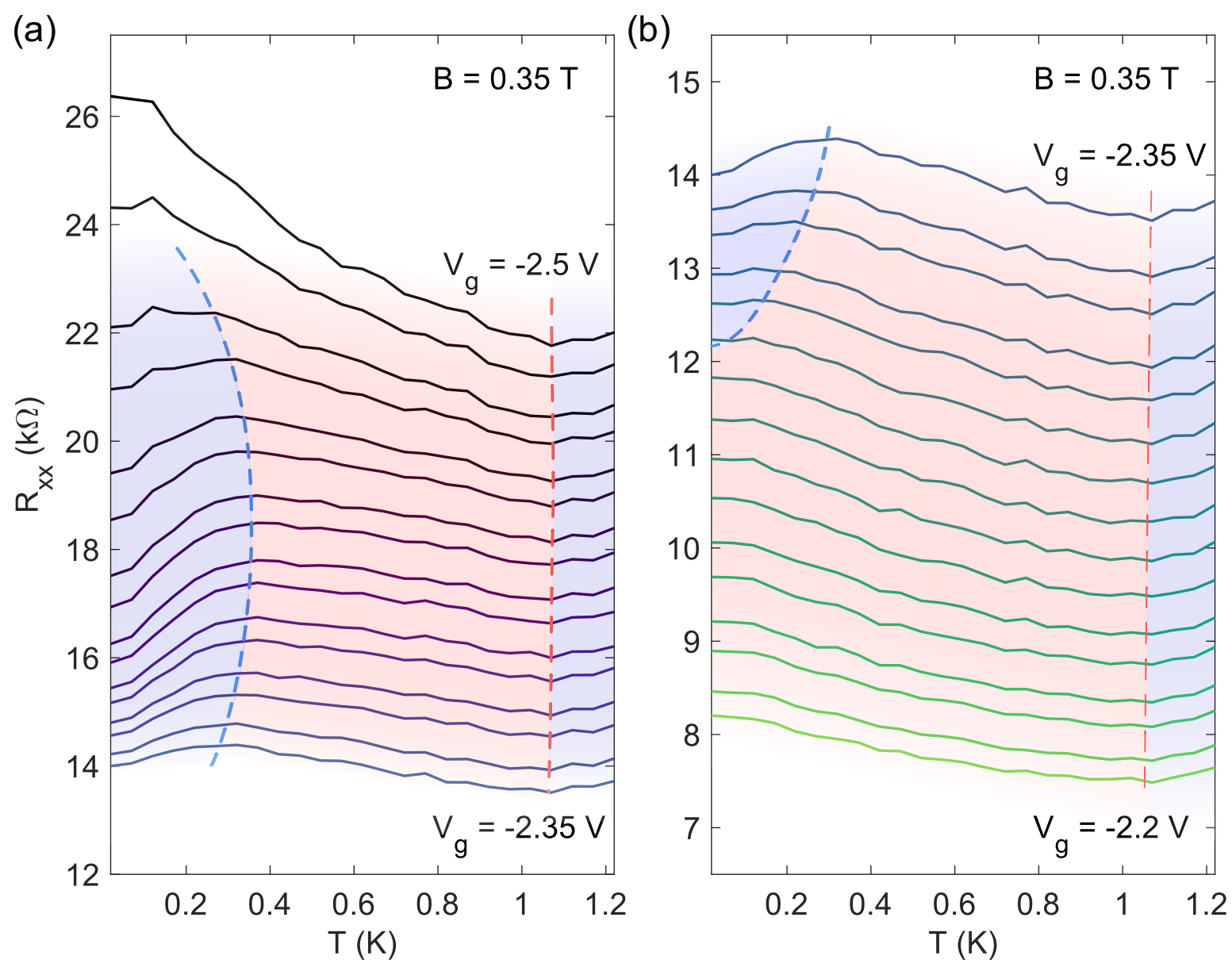

**Figure 4. Re-entrance of the electron Wigner solid.** Shown are resistivity traces near the conduction band edge at a magnetic field ($B$) of 0.35 T at different gate voltages ($V_g$). The traces are recorded between $V_g$ = -2.5 V and $V_g$ = -2.35 V in (a) and between $V_g$ = -2.35 V and $V_g$ = -2.2 V in (b). The $V_g$ increments are 0.01 V in both graphs. The dashed lines indicate the boundaries between the solid and liquid phases.

Importantly, there are additional transitions between electron solids and liquids in small magnetic fields as the temperature is increased. As highlighted by the box in Fig. 2(c), the low-temperature liquid phase that emerged from the solid near $\nu = 1$ transitions back to a solid at ~ 400 mK, visible as the red region that is marked by a green circle. The finite temperature transition is shown in more detail in Fig. 4 for $B$ = 0.35 T, which shows line traces of $R_{xx}$ taken at different values of $V_g$ near the vicinity of the liquid phase. The dashed lines indicate the boundaries between the liquid ($dR_{xx}/dT > 0$, blue shaded) and solids ($dR_{xx}/dT < 0$, red shaded) regimes. The red dashed lines indicate the transition from the quantum Hall liquid to the electron solid (stable at intermediate temperatures) while the blue dashed lines indicate the melting of the Wigner solid. In Supplementary Figure 6, we compare AC and DC resistivity data near the transition. Linear *I*-*V* behavior is seen in both AC and DC measurements and the sign change in $dR_{xx}/dT$ is observed at the same temperature. The linear *I*-*V* behavior of the solid and its relatively low resistivity is expected due to the absence of pinning in this temperature regime and it was also observed in the other stability regimes of the Wigner solids in which there is no competition with the Hall liquid. Importantly, these regimes are smoothly connected in behavior to the reentrant solid (see also Supplementary Figure 3), providing additional support to the interpretation that the reentrant phase is the Wigner solid, and not a new phase. We can also rule out a disorder-driven insulator, given that the transition to a metallic phase at lower temperature, which would not be expected for a disorder-driven transition.

Interestingly, there is no reentrant insulator or solid on the hole side, in sharp contrast to the electron side. The Hall liquid corresponding to $\nu = -1$ emerges at around 3.3 T and the transition is almost temperature independent (see Fig. 3). Near $\nu = -1$, $dR_{\text{xx}}/dT$ is monotonic and approaches zero at the critical field strength.

## Discussion

We begin with discussing the competition between the Wigner solids and the quantum Hall liquids in finite fields near integer filling factors *in the low temperature limit*. Our results can be elegantly explained by an internal energy-driven competition, as suggested recently by Reddy and Fu [24]. The internal energies per particle of the solid ($E_{\text{S}}$) and the liquid ($E_{\text{L}}$) depend on the magnetic field and the carrier density as shown schematically in Fig. 5.

Due to Landau quantization, $E_{\text{L}}$ exhibits local minima whenever the carrier density corresponds to an integer filling factor, where the bulk becomes incompressible. The depth of the minima in $E_{\text{L}}$ scales with the magnitude of the cyclotron gap (i.e., magnetic field). In contrast, $E_{\text{S}}$ increases monotonically with magnetic field [24]. The result is a critical field $B_{\text{c}}$, above which $E_{\text{S}} > E_{\text{L}}$, i.e., the liquid becomes the stable ground state at $T = 0$ K. We extend the model by Reddy and Fu to include the competition as a function of carrier density for massive Dirac fermions, as shown in Fig. 5(b). On the high carrier density side of the phase stability range of the solid, $E_{\text{S}}$ increases with carrier density and, above a critical density, the solid will melt into a quantum liquid [25,27,28]. Below this critical carrier density, $E_{\text{L}}$ can drop below $E_{\text{S}}$ at integer values of ν, causing the emerging quantum Hall liquid to consume the solid. In finite magnetic fields and at low temperatures, the liquid and solid phases can alternate with carrier density. Figure 5(b) thus explains the experimental observations of alternating regions of $dR_{xx}/dT > 0$ and $dR_{xx}/dT < 0$ seen in Fig. 3. At large carrier densities, $E_{\text{S}} > E_{\text{L}}$ everywhere and there are no Wigner solids.

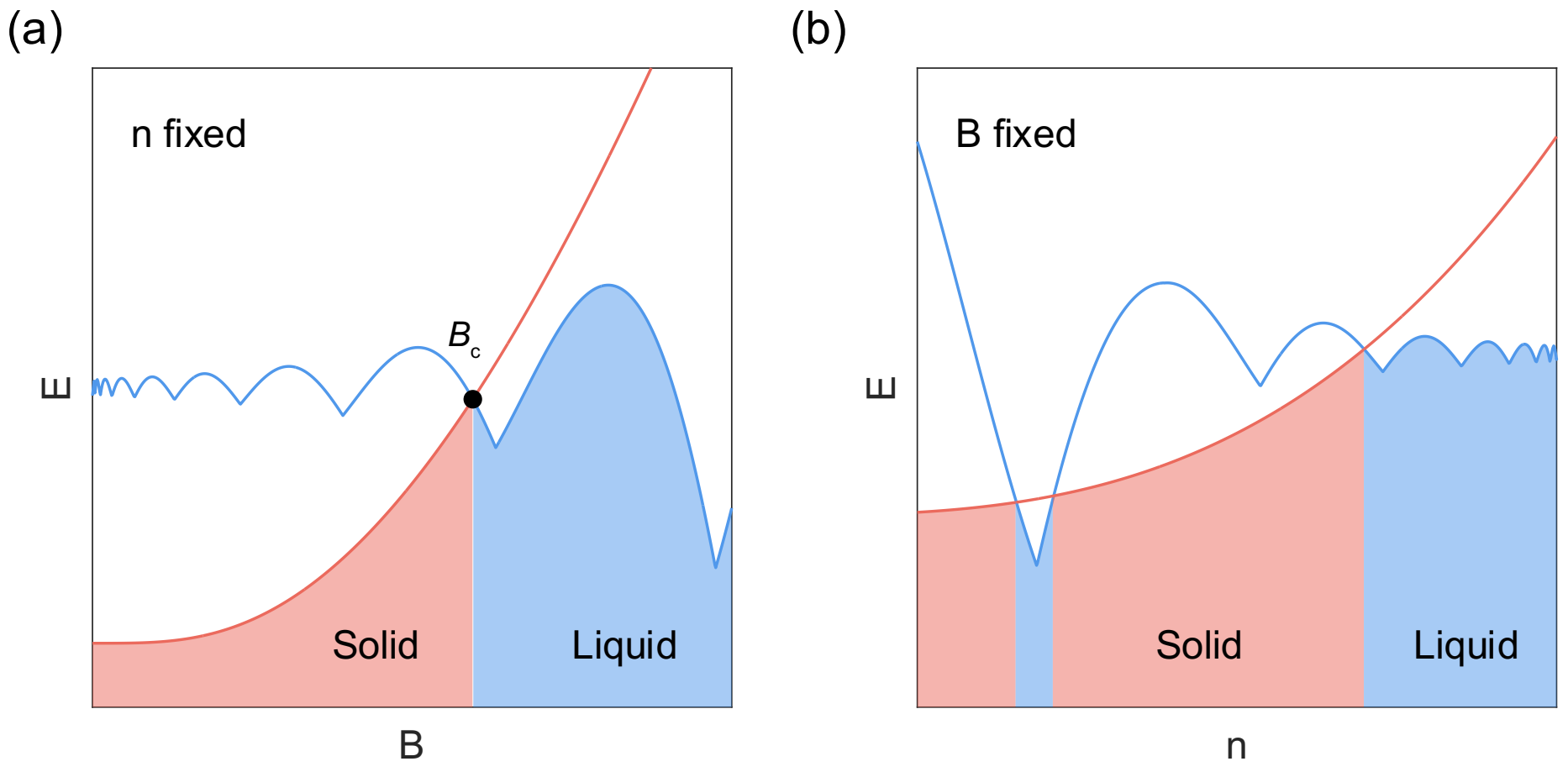

**Figure 5. Internal energy (*E*) competition between Wigner solids and quantum Hall liquids**. Shown are the energies of the liquid and solid ($E_{\mathrm{L}}$ and $E_{\mathrm{S}}$ in the text) as a function of (a) magnetic field ($B$) and (b) carrier density ($n$). Panel (a) is similar to the one shown in Reddy and Fu [24]. The color of the shaded regions indicates the stable phase. The depth of the minima in $E_{\mathrm{L}}$ scales with the cyclotron gap.

We next turn to the remarkable non-monotonic $dR_{xx}/dT$ behavior seen at *finite temperatures* and fields near $\nu = 1$ (Figs. 2 and 4). As discussed above, from the experimental data, near $B = 0.35$ T, $E_{\mathrm{S}} \approx E_{\mathrm{L}}$. Thus, at this field, the reentrant electron solid at intermediate temperatures strongly suggests that we need to consider the entropy term in the free energy in the competition between the liquid and solid. In general, it is known that the spin degrees of freedom are the largest source of entropy in the Wigner solids [29]. Each spin contributes an entropy of $k_B \ln(2)$, causing the entropy contribution to the free energy of the solid ($F_S$) to decrease linearly with temperature. With increasing temperature, the Wigner solid will eventually melt. In contrast, the liquid has negligible spin entropy, due to the Pauli exclusion principle [29]. Theory, therefore, predicts a Pomeranchuk effect (re-entrant solid) at finite temperatures in zero magnetic fields [25,28,29].

Recall, however, that here we observe a re-entrant solid *at finite fields*, while the traditional Pomeranchuk effect only considers zero fields. Applying an out-of-plane field quenches the spin degrees of freedom of the solid and may open a Zeeman gap. As depicted in Fig. 6(a), in this case, $F_S$ depends on the thermal excitation across the Zeeman gap ($\Delta_Z$). To estimate $\Delta_Z$ at 0.35 T, we use the $g$ factor calculated in ref. [30] from an effective $k \cdot p$ model for $Cd_3As_2$ films of this thickness ($g$ ~ 10), which gives $\Delta_Z = 0.2$ meV. The cyclotron energy is calculated to be about $\Delta_C = 2$ meV from the same $k \cdot p$ model. We estimate the entropy contribution to the free energy per particle in the $\nu = 1$ liquid as $F_L = -2k_BT\ln(1 + e^{-\Delta_C/2k_BT})$, assuming two Landau levels separated by $\Delta_C$ with the chemical potential situated halfway between them. We note that in $Cd_3As_2$ quantum wells, the $\nu = 1$ plateau appears immediately because of spin-orbit coupling, which causes a large orbital gap between the spin-polarized $n = 0$ and $n = 1$ Landau levels (see, e.g., ref. [31]) and it is not a many-body effect. We approximate $F_S$ by a canonical ensemble two-level system, $F_S = -k_BT\ln(1 + e^{-\Delta_Z/k_BT})$. Figure 6(b) shows $F_S$ and as $F_L$ a function of temperature for a magnetic field of 0.35 T. The solid green line is the free-spin estimate (traditional zero-field Pomeranchuk effect). The large $\Delta_C$ causes $F_L$ to be almost independent of temperature (purple line). For $F_S$, we use a range of $g$ factors ($5 \leq g \leq 15$) to estimate $\Delta_Z$, recognizing that $g$ strongly depends on various microscopic parameters of the quantum well (bulk $g$-factors are very large in $Cd_3As_2$ [32]). Moreover, in a Wigner solid, a Bloch wave description no longer holds and the large separation of wave packets could result in smaller $g$ factors. Importantly, Fig. 6(b) shows that in the experimentally relevant temperature regime from 0.4 K to 1 K, where we observe the reentrant solid, $F_S < F_L$, as a result of spin entropy.

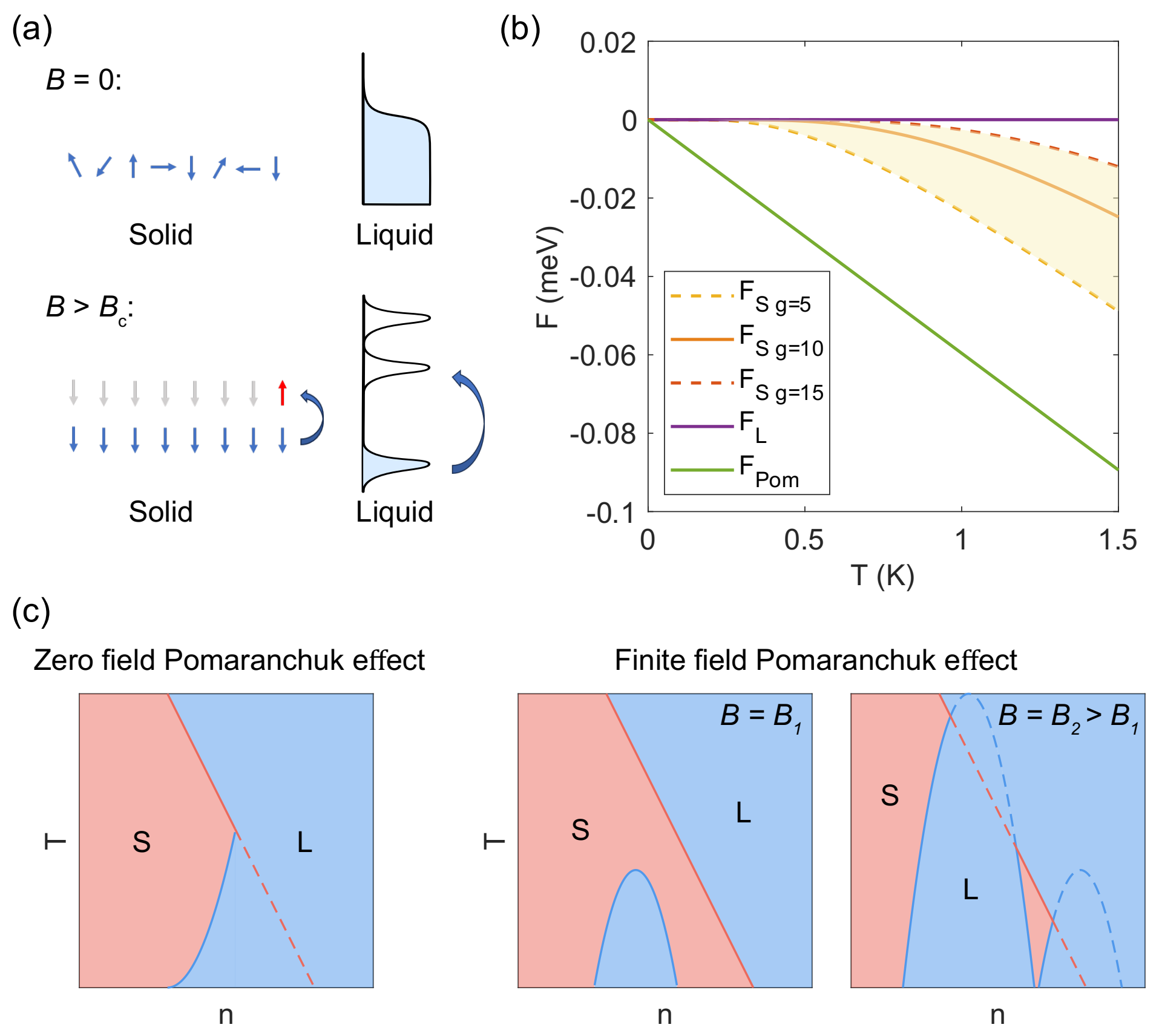

**Figure 6. Role of spin entropy in the competition between Wigner solids and quantum Hall liquids.** (a) Spins in Wigner solids and Fermi liquids. Upper panel: at zero field ($B$), the spins in the Wigner solids are free to rotate, maximizing the spin entropy. As a result, at finite temperatures, $F_S < F_L$ ($F_S$ and $F_L$ are the entropy contributions to the free energy of the solid and liquid, respectively). Lower panel: Wigner solids and liquid in $B > B_c$ ($B_c$ is defined in Fig. 5(a)). Cyclotron and Zeeman gaps are induced in the energy spectra of liquid and solid, respectively. (b) Entropy contribution ($F$) to the free energy as a function of temperature (see text). At elevated temperatures, the Wigner solid can gain a free energy advantage, which depends on the magnitude to the Zeeman gap. The subscript indicates the $g$-factor. (c) Schematic phase diagrams for different values of the magnetic field. Here, $n$ is the carrier density, $T$ is the temperature, S and L indicate the solid and liquid, respectively.

The same free energy consideration neatly explains the *absence* of a re-entrant solid on the hole side. For hole carriers, the quantum Hall liquid corresponding to $\nu = -1$ emerges only at large fields (> 3 T). At such high fields, both $\Delta_C$ and $\Delta_Z$ are too large for the spin entropy term to play a significant role. This also explains why the Pomeranchuk effect has not been seen in studies of Wigner solids that were carried out at high magnetic fields.

For spin entropy to dominate, the inter-electron spacing is an important parameter. The reentrant solid appears at a carrier density $n_s \sim 2 \times 10^{10}$ cm$^{-2}$ (see Supplementary Figure 1a), which yields a Wigner-Seitz radius of $a = \frac{1}{\sqrt{\pi n_s}} = 40$ nm. The second important radius is the effective Bohr radius, $a_B^* = \frac{\epsilon_r}{m^*/m_e} a_B$, for which we only have rough estimate. Assuming a relative permittivity $\epsilon_r = 20$ and an electron effective mass $m^* \sim 0.04 m_e$ (determined from SdH oscillations at $n_s \sim 5 \times 10^{11}$ cm$^{-2}$) yields $a_B^* = 26$ nm, i.e., the carriers are reasonably well separated. Moreover, the $m^*$ at the bottom of the conduction band is likely significantly larger, given that it increases sharply with decreasing carrier density. The effective mass of the holes is much higher, but we do not have a reliable value.

Figure 6(c) summarizes the picture of the Wigner solid-quantum Hall liquid transitions. A phase diagram depicting the conventional, zero-field Pomaranchuk effect is shown in the left panel. At elevated temperatures where spin entropy contributes to the physics, the main consequence of the Pomeranchuk effect is to shift the critical point for melting to lower carrier densities. In contrast, the combination of the finite-field Pomeranchuk effect and the internal energy competition of the zero-temperature ground states results in a much richer phase diagram. At low carrier densities and in zero magnetic field, $E_{\mathrm{S}}$ is lower than that of a Fermi liquid (or gas) and Wigner solids form near the conduction and valence band edges. At low temperatures and in magnetic fields, $E_{\mathrm{L}}$ is an oscillatory function of the carrier density (see Fig. 5). Near $\nu = 1$, the energy of the emerging quantum Hall liquid drops below that of the solid, $E_{\mathrm{L}}(B > 0) < E_{\mathrm{S}}(B > 0)$, causing the Wigner solid to be consumed, as shown in the middle panel. As the temperature is raised, however, and as long as $\Delta_{\mathrm{Z}}$ remains sufficiently small, the entropic contribution can begin to dominate ($F_S < F_L$), driving the reentrance of the solid phase before it eventually thermally melts at higher temperatures (middle panel). The result closely mimics the zero-field Pomeranchuk effect. With increasing magnetic field, the stability range of the quantum Hall liquid expands in parameter space. In addition, the melting temperature of the Wigner solid decreases (right panel). This gives rise to itinerant liquid (or gas) phase around 1 K as seen in Fig. 4.

Finally, we note that in addition to electron spin, there could also be other sources of entropy that favor the solid at high temperature. For example, theoretical proposals predict stripes and bubble phases at intermediate temperatures [25], which could also provide configurational entropy. We leave the study of the higher temperature phases, such as the melting into two-phase regions or microemulsions [20,25,33], and their role in the transport behavior to future work.

To conclude, using transport studies of Wigner solid to quantum Hall liquid phase transitions, our study supports a recently proposed zero-temperature energetic model that explains the surprising destabilization of zero-field Wigner solids by integer quantum Hall liquids in magnetic fields. Moreover, our data provides evidence of a remarkable entropy-driven reentrance of the solid phase at finite temperatures and small fields that is critical in explaining the finite-temperature competition between Wigner solids and quantum Hall phase. The physics uncovered here resembles the free-spin Pomeranchuk effect and adds to a growing body of experimental work of carrier solidification upon heating in fermionic systems [12,33-36]. The Pomeranchuk effect was observable because of the fact that these Wigner solids can exist at low fields, where the Zeeman gap is not too large. Zero-field Wigner solids thus offer new opportunities to elucidate the interplay between a topological and correlated electron states.

## METHODS

### Device Fabrication

Measurements were carried out on top-gated Hall bars fabricated from high-mobility (001) $Cd_3As_2$ films, grown by molecular beam epitaxy on a lattice-matched $Al_{0.45}In_{0.55}Sb$ buffer layer on a (001)-oriented GaSb substrate, as described in detail elsewhere [37,38]. This particular quantum well had a thickness that was close to the critical thickness of transition from the two-dimensional topological insulator phase to trivial insulator (see refs. [23,39,40] for a discussion of the thickness dependence). Applying standard photolithography and lift-off techniques, a Hall bar was defined with mesa widths and lengths of 25 μm and 250 μm, respectively. Details of the device fabrication can be found in the Supplementary Information and in prior publications [23,41]. Carrier densities and Hall mobilities of the film are shown in the Supplementary Information.

## Measurement

Electrical transport measurements were carried out in an Oxford Triton dilution refrigerator system with a base temperature of 12 mK. Two-dimensional resistivities and conductivities were obtained using a four-probe AC lock-in technique with frequency of ~ 2 Hz and an excitation current of 2 nA. A constant applied current was achieved by combining a lock-in amplifier and a voltage controlled current source. The temperature was varied in increments of 50 mK, the magnetic field increments were 0.2 T, and $V_g$ was swept at a rate of 0.02 V/s. For DC bias sweeps, the sweep rate was 1 mV/s. Due to the high impedance inherent to Wigner solid states, a very low frequency was used to reliably measure the longitudinal resistivity. Longitudinal ($\sigma_{xx}$) and transverse conductivities ($\sigma_{xy}$) were calculated by tensor inversion, i.e., $\sigma_{xx} = R_{xx}/\left(R_{xx}^2 + R_{xy}^2\right)$ and $\sigma_{xy} = R_{xy}/\left(R_{xx}^2 + R_{xy}^2\right)$. All measurements reported here were carried out on lines that were unfiltered. The temperatures in the main text refer to the set bath temperature and the actual electron temperature is likely higher, especially below a bath temperature of 100 mK. The higher temperature causes, for example, the onset of sliding behavior to appear less abrupt than in ref.

[11], which was carried on filtered lines (see ref. [11] for DC *I-V* traces as a function of temperature). Nonetheless, even without filtering, we still observed resistivity changes with temperature between 20 mK and 100 mK, allowing for the study of the temperature dependent behavior to below 100 mK.

## DATA AVAILABILITY

The datasets generated and/or analyzed during the current study are not publicly available due to concerns by the authors about the data being used without proper attribution by AI agents, but are available to other humans from the corresponding author on reasonable request.

## Acknowledgments

The authors thank Junkai Dong and Wangqian Miao for many useful discussions.

**Funding:** The work was supported by the Gordon and Betty Moore Foundation's EPiQS Initiative, Grant GBMF11944, and a MURI program of the Air Force Office of Scientific Research under grant no. FA9550-22-1-0270. Partial funding was also provided by a grant from the Army Research Office (W911-NF-23-1-0371).

## Author Contributions

B. L. performed all measurements and analyzed the data. S. M. assisted with measurements. S. A. grew the film. E.B. assisted with the theoretical interpretation. B.L. and S. S. prepared the manuscript and all authors commented on it. **Competing Interests**: The authors declare no competing financial or non-financial interests.

## ADDITIONAL INFORMATION

**Supplementary information:** The online version contains supplementary material available at (link by publisher).

## References

[1] E. Wigner, *On the Interaction of Electrons in Metals*, Phys. Rev. **46**, 1002–1011 (1934).

[2] E. Y. Andrei, G. Deville, D. C. Glattli, F. I. B. Williams, E. Paris, and B. Etienne, *Observation of a Magnetically Induced Wigner Solid*, Phys. Rev. Lett. **60**, 2765 (1988).

[3] H. W. Jiang, R. L. Willett, H. L. Stormer, D. C. Tsui, L. N. Pfeiffer, and K. W. West, *Quantum liquid versus electron solid around ν=1/5 Landau-level filling*, Phys. Rev. Lett. **65**, 633-636 (1990).

[4] V. J. Goldman, M. Santos, M. Shayegan, and J. E. Cunningham, *Evidence for two-dimentional quantum Wigner crystal*, Phys. Rev. Lett. **65**, 2189 (1990).

[5] Y.-C. Tsui, M. He, Y. Hu, E. Lake, T. Wang, K. Watanabe, T. Taniguchi, M. P. Zaletel, and A. Yazdani, *Direct observation of a magnetic-field-induced Wigner crystal*, Nature **628**, 287–292 (2024).

[6] J. Yoon, C. C. Li, D. Shahar, D. C. Tsui, and M. Shayegan, *Wigner Crystallization and Metal-Insulator Transition of Two-Dimensional Holes in GaAs at B=0*, Phys. Rev. Lett. **82**, 1744 (1999).

[7] S. V. Kravchenko, J. A. A. J. Perenboom, and V. M. Pudalov, *Two-dimensional electron solid formation in Si inversion layers*, Phys. Rev. B **44**, 13513 (1991).

[8] T. Smoleński, P. E. Dolgirev, C. Kuhlenkamp, A. Popert, Y. Shimazaki, P. Back, X. Lu, M. Kroner, K. Watanabe, T. Taniguchi, I. Esterlis, E. Demler, and A. Imamoğlu, *Signatures of Wigner crystal of electrons in a monolayer semiconductor*, Nature **595**, 53–57 (2021).

[9] M. S. Hossain, M. K. Ma, K. A. V. Rosales, Y. J. Chung, L. N. Pfeiffer, K. W. West, K. W. Baldwin, and M. Shayegan, *Observation of spontaneous ferromagnetism in a two-dimensional electron system*, Proc. Nat. Acad. Sci. **117**, 32244-32250 (2020).

[10] J. Falson, I. Sodemann, B. Skinner, D. Tabrea, Y. Kozuka, A. Tsukazaki, M. Kawasaki, K. von Klitzing, and J. H. Smet, *Competing correlated states around the zero-field Wigner crystallization transition of electrons in two dimensions*, Nat. Mater. **21**, 311 (2022).

[11] S. Munyan, S. Ahadi, B. H. Guo, A. Rashidi, and S. Stemmer, *Evidence of Zero-Field Wigner Solids in Ultrathin Films of Cadmium Arsenide*, Phys. Rev. X **14**, 041037 (2024).

[12] T. Han, J. P. Butler, S. Ye, Z. Hua, S. Dutta, Z. Hadjri, Z. Wu, J. Yang, J. Seo, P. Pattanakanvijit, E. Aitken, K. Watanabe, T. Taniguchi, P. Xiong, E. Zeldov, Z. Lu, R. Ashoori, and L. Ju, *Evidence of Metallic Wigner Crystal in Rhombohedral Graphene*, arXiv:2604.00113 [cond-mat.mes-hall] (2026).

[13] A. M. Seiler, M. Statz, C. Eckel, I. Weimer, J. Pöhls, K. Watanabe, T. Taniguchi, F. Zhang, and R. T. Weitz, *Signatures of sliding Wigner crystals in bilayer graphene at zero and finite magnetic fields*, Nat. Comm. **16**, 8921 (2025).

[14] P. Brussarski, S. Li, S. V. Kravchenko, A. A. Shashkin, and M. P. Sarachik, *Transport evidence for a sliding two-dimensional quantum electron solid*, Nat. Comm. **9**, 3803 (2018).

[15] M. S. Hossain, M. K. Ma, K. A. Villegas-Rosales, Y. J. Chung, L. N. Pfeiffer, K. W. West, K. W. Baldwin, and M. Shayegan, *Anisotropic Two-Dimensional Disordered Wigner Solid*, Phys. Rev. Lett. (accepted) **129**, 036601 (2022).

[16] T. Knighton, J. H. Zhe Wu, A. Serafin, J. S. Xia, L. N. Pfeiffer, and K. W. West, *Evidence of two-stage melting of Wigner solids*, Phys. Rev. B **97**, 085135 (2018).

[17] H. Deng, L. N. Pfeiffer, K. W. West, K. W. Baldwin, L. W. Engel, and M. Shayegan, *Probing the Melting of a Two-Dimensional Quantum Wigner Crystal via its Screening Efficiency*, Phys. Rev. Lett. **122**, 116601 (2019).

[18] S. Chen, R. Ribeiro-Palau, K. Yang, K. Watanabe, T. Taniguchi, J. Hone, M. O. Goerbig, and C. R. Dean, *Competing Fractional Quantum Hall and Electron Solid Phases in Graphene*, Phys. Rev. Lett. **122**, 026802 (2019).

[19] V. Shingla, S. A. Myers, L. N. Pfeiffer, K. W. Baldwin, and G. A. Csáthy, *Particle-hole symmetry and the reentrant integer quantum Hall Wigner solid*, Comm. Phys. **4**, 204 (2021).

[20] B. Spivak and S. A. Kivelson, *Transport in two dimensional electronic micro-emulsions*, Ann. Phys. **321**, 2071-2115 (2006).

[21] N. Deng, A. Kumar, M. J. Manfra, L. N. Pfeiffer, K. W. West, and G. A. Csáthy, *Collective Nature of the Reentrant Integer Quantum Hall States in the Second Landau Level*, Phys. Rev. Lett. **108**, 086803 (2012).

[22] K. A. V. Rosales, S. K. Singh, H. Deng, Y. J. Chung, L. N. Pfeiffer, K. W. West, K. W. Baldwin, and M. Shayegan, *Melting phase diagram of bubble phases in high Landau levels*, Phys. Rev. B **104**, L121110 (2021).

[23] A. C. Lygo, B. Guo, A. Rashidi, V. Huang, P. Cuadros-Romero, and S. Stemmer, *Two-dimensional topological insulator state in cadmium arsenide thin films*, Phys. Rev. Lett. **130**, 046201 (2023).

[24] A. P. Reddy and L. Fu, *Quantum melting a Wigner crystal into Hall liquids*, Phys. Rev. B **113**, L161403 (2026).

[25] B. Spivak and S. A. Kivelson, *Phases intermediate between a two-dimensional electron liquid and Wigner crystal*, Phys. Rev. B **70**, 155114 (2004).

[26] D. G. Polyakov and B. I. Shklovskii, *Activated Conductivity in the Quantum Hall Effect*, Physical Review Letters **73**, 1150-1153 (1994).

[27] P. M. Platzman and H. Fukuyama, *Phase diagram of the two-dimensional electron liquid*, Phys. Rev. B **10**, 3150 (1974).

[28] B. K. Clark, M. Casula, and D. M. Ceperley, *Hexatic and Mesoscopic Phases in a 2D Quantum Coulomb System*, Phys. Rev. Lett. **103**, 055701 (2009).

[29] B. Spivak, *Phase separation in the two-dimensional electron liquid in MOSFET's*, Phys. Rev. B **67**, 125205 (2003).

[30] W. Miao, B. Guo, S. Stemmer, and X. Dai, *Engineering the in-plane anomalous Hall effect in $Cd_3As_2$ thin films*, Phys. Rev. B **109**, 155408 (2024).

[31] B. Liu, W. Miao, S. Munyan, S. Stemmer, and X. Dai, *Zeroth Landau level crossings in two-dimensional topological insulators*, Phys. Rev. B **111**, 195136 (2025).

[32] F. A. P. Blom, J. W. Cremers, J. J. Neve, and M. J. Gelten, *Anisotropy of the Electronic g*-Factor in Cadmium Arsenide*, Solid State Comm. **33**, 69-73 (1980).

[33] J. Sung, J. Wang, I. Esterlis, P. A. Volkov, G. Scuri, Y. Zhou, E. Brutschea, T. Taniguchi, K. Watanabe, Y. Yang, M. A. Morales, S. Zhang, A. J. Millis, M. D. Lukin, P. Kim, E. Demler, and H. Park, *An electronic microemulsion phase emerging from a quantum crystal-to-liquid transition*, Nat. Phys. **21**, 437–443 (2025).

[34] A. Rozen, J. M. Park, U. Zondiner, Y. Cao, D. Rodan-Legrain, T. Taniguchi, K. Watanabe, Y. Oreg, A. Stern, E. Berg, P. Jarillo-Herrero, and S. Ilani, *Entropic evidence for a Pomeranchuk effect in magic-angle graphene*, Nature **592**, 214–219 (2021).

[35] N. Matsuyama, S. Yokomori, T. Nomura, Y. Ishii, H. Hayashi, H. Ishikawa, K. Matsui, H. Mori, K. Kindo, Y. H. Matsuda, and S. Imajo, *Fate of Pomeranchuk effect in ultrahigh magnetic fields*, Nat. Comm. **17**, 367 (2026).

[36] Y. Saito, F. Yang, J. Ge, X. Liu, T. Taniguchi, K. Watanabe, J. I. A. Li, E. Berg, and A. F. Young, *Isospin Pomeranchuk effect in twisted bilayer graphene*, Nature **592**, 220–224 (2021).

[37] M. Goyal, S. Salmani-Rezaie, T. N. Pardue, B. H. Guo, D. A. Kealhofer, and S. Stemmer, *Carrier mobilities of (001) cadmium arsenide films*, APL Mater. **8**, 051106 (2020).

[38] B. H. Guo, A. C. Lygo, T. N. Pardue, and S. Stemmer, *Hall bar measurements of topological surface states of (001) cadmium arsenide thin films interfaced with superconductors*, Phys. Rev. Mater. **6**, 034203 (2022).

[39] Z. J. Wang, H. M. Weng, Q. S. Wu, X. Dai, and Z. Fang, *Three-dimensional Dirac semimetal and quantum transport in* $Cd_3As_2$, Phys. Rev. B **88**, 125427 (2013).

[40] B. Guo, W. Miao, V. Huang, A. C. Lygo, X. Dai, and S. Stemmer, *Zeeman Field-Induced Two-Dimensional Weyl Semimetal Phase in Cadmium Arsenide*, Phys. Rev. Lett. **131**, 046601 (2023).

[41] B. Guo, A. C. Lygo, X. Dai, and S. Stemmer, $\nu = 0$ *quantum Hall state in a cadmium arsenide thin film*, APL Mater. **10**, 091116 (2022).

## Supplementary Information

# Quantum melting and Pomeranchuk effect of zero-field Wigner solids

**Boliang Liu[1], Simon Munyan[1], Sina Ahadi[1], Erez Berg[1], and Susanne Stemmer[1]**
[1]Materials Department, University of California, Santa Barbara, California 93106-5050, USA

### Device Fabrication

The Hall bar was fabricated using standard photolithography techniques. The mesa (width of 25 µm and length of 250 µm) was wet etched using a solution of $NH_4OH$, $H_2O_2$ and $H_2O$. Ohmic pads were deposited via electron beam deposition of Ti/Pt/Au (10 nm/20 nm/200 nm). Devices were treated with a low-power oxygen plasma prior to the deposition of a blanket $Al_2O_3$ gate dielectric via atomic layer deposition at 120 °C. The dielectric was etched from the ohmic pads with dilute tetramethylammonium hydroxide. Lastly, a gate electrode of Ni/Au (5 nm/150 nm) was deposited via thermal evaporation.

### Transverse Conductivity Data

In Supplementary Figure 1, we show the measured carrier densities ($n$), Hall mobilities ($\mu$), and transverse conductivity ($\sigma_{xy}$) as a function of the gate voltage ($V_g$) and the magnetic field ($B$) in the range where they could be reliably determined (e.g., no longitudinal-transverse signal mixing, as is found in the charge neutrality gap). An optical micrograph of the device measured is also provided.

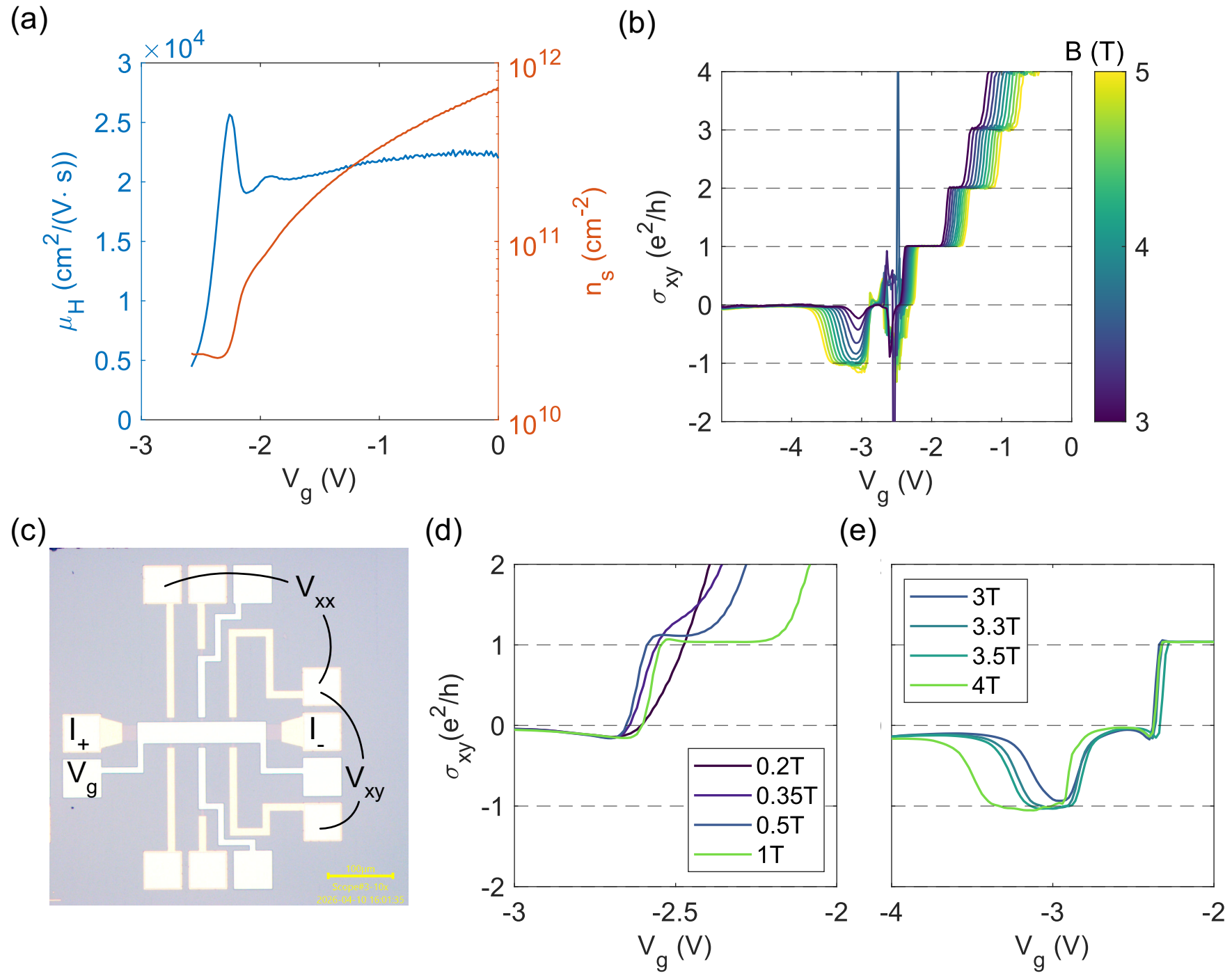

**Supplementary Figure 1:** (a) Carrier density ($n$) and Hall mobility ($\mu$) for the conduction subband. The carrier density is as low as $2\times10^{10}$ cm$^{-2}$ near the conduction band edge, which corresponds to an average carrier spacing of 70 nm. (b) Transverse conductivity $\sigma_{xy}$ as a function of magnetic field $B$ and of gate voltage $V_g$. (c) A micrograph of the Hall bar device. $I^+$ and $I^-$ indicate the source and drain in both AC and DC measurements. The top gate voltage is $V_g$. (d) $\sigma_{xy}$ traces near the conduction band edge at field ranges near where electron-like Wigner solids transitions into the quantum Hall plateau $\nu = 1$. (e) $\sigma_{xy}$ traces near the valence band edge in field ranges near where the hole-like Wigner solids transitions into quantum Hall plateau $\nu = -1$.

## Complete Set of DC Current-Voltage (*I-V*) Traces

In Supplementary Figure 2, we show DC *I-V* traces over a wide range of gate voltages ($V_g$). The step size in $V_g$ values is 0.1 V. Panel (a) covers $V_g$ values from $V_g$ = -2.4 V to $V_g$ = - 3 V, which includes the electron Wigner solid. Hysteretic *I-V* traces and sharp threshold voltages are visible. As the carrier density ($V_g$) increases, the hysteresis gradually vanishes and the *I-V* relationship becomes more linear. This behavior implies the quantum melting of the pinned Wigner solid to a de-pinned Wigner solid and eventually to a Fermi liquid. Supplementary Figure 2(b) shows *I-V* traces from $V_g$ = -2.7 V to $V_g$ = - 4.6 V. The measured resistance is the largest between $V_g$ = -3 V to $V_g$ = - 4 V, which includes the charge neutrality gap and the hole Wigner solid.

When the bulk is too resistive, charging effects dominate over the signal from the depinning of the Wigner solid. A high resistivity effectively turns the Hall bar into a tunnel junction with a barrier. When the top gate tunes the chemical potential into the barrier, applying a DC bias across the device changes the local chemical potential and accumulates charge. This is illustrated in Fig. Supplementary Figure 2(c), where the green, blue, and orange color blocks represent the combined Wigner solid-charge neutrality gap in energy. A voltage bias ($V_{DC}$) is applied from to the right contact. The dashed line segment on the right of the gap represents the chemical potential set by top gate before applying $V_{DC}$. Applying a DC voltage effectively moves the chemical potential up or down, depending of the sign of $V_{DC}$.

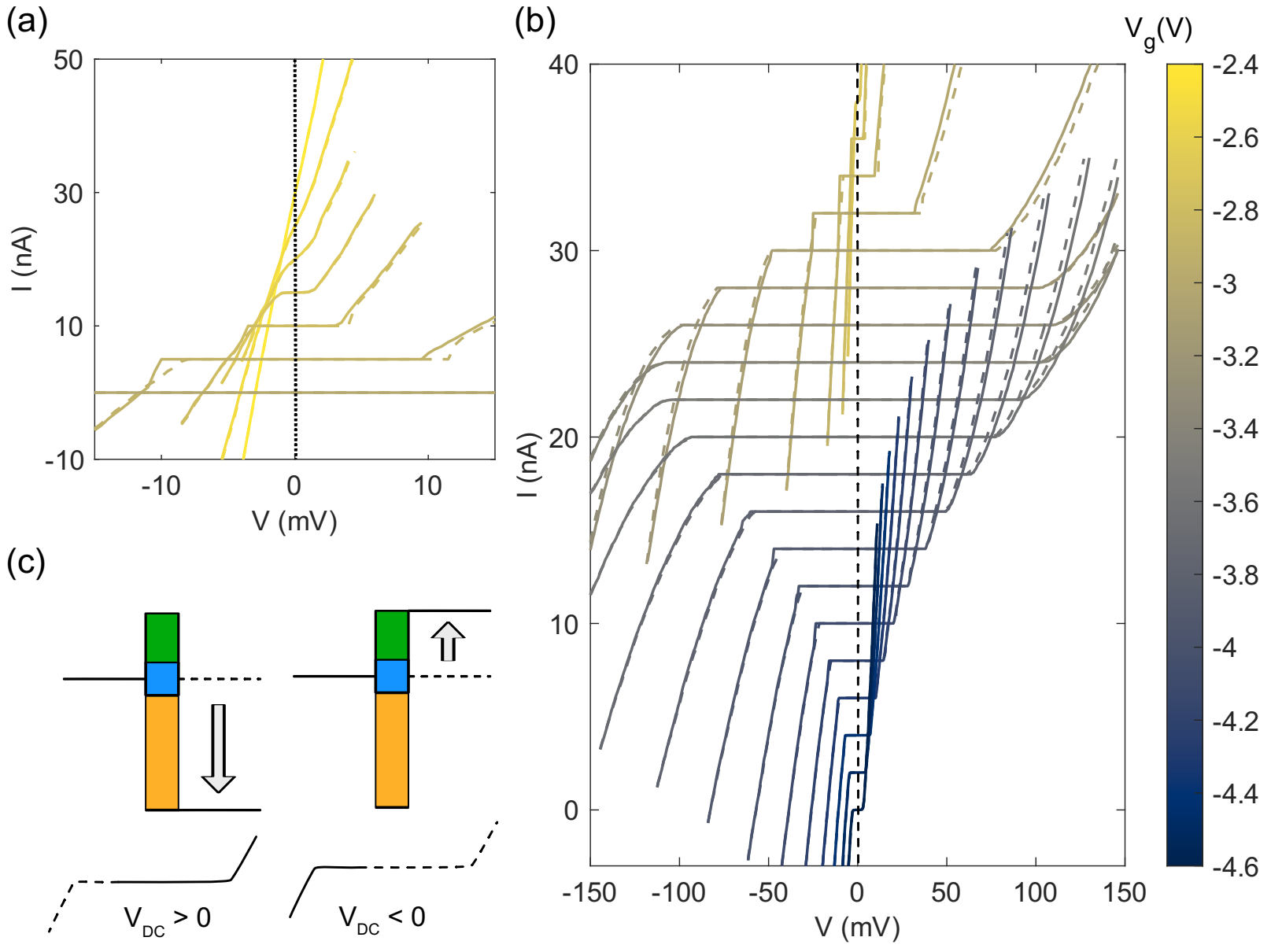

**Supplementary Figure 2:** (a-b) DC *I-V* traces taken at different gate voltages using a step size of $\Delta V_g$ of 0.1 V. (a) The top trace is taken at $V_g$ = -2.4 V and the bottom trace is $V_g$ = - 3 V. Each trace is manually offset by 5 nA from the next. (b) The top trace is taken $V_g$ = -2.7 V and the bottom trace is taken at $V_g$ = - 4.6. Each trace is manually offset by 2 nA from the next. (c) Schematics showing the charging effect when the sample is highly resistive. The green, blue, and orange color blocks represent the Wigner solids and charge neutrality gaps.

## Temperature Dependences of the AC resistivity and DC *I-V* Traces

In Supplementary Figures 3 and 4, we show AC resistivity ($R_{xx}$) traces at different temperatures ($T$), from which the values of $dR_{xx}/dT$ shown in the main text were calculated. Temperatures range from 20 mK to 1.22 K with a step size of 50 mK so that $dR_{xx}/dT = [R(T_2) - R(T_1)]/0.05\ \Omega/\text{K}$. The resistivity is related to measured resistance by the aspect ratio of 1:4. In Supplementary Figures 3(a), data points corresponding to $R_{xx} > 0.7$ MΩ were removed due to the presence of artifacts. These artifacts are due to the large resistance of the device in the insulating regime that outcompetes the input impedance of lock-in amplifiers. Insulating (Wigner) and emerging quantum Hall liquid regimes are identified by their temperature dependence (see main text).

Supplementary Figure 5(a) shows DC *I-V* traces as a function of temperature ($T$). The gate voltage is set to $V_g$ = -3.6 V. The sharp threshold voltage behavior and the non-linear *I-V* diminish as $T$ increases. Nonetheless, the solid shows activated behavior to 1.2 K. We calculate the total resistance $R = V_{DC}/I_c$ at a constant DC current ($I_c$). This can also be viewed as the resistance required to drive the electron solid to a certain speed. In Supplementary Figure 5(b) we showed logarithmic resistance ($\ln R$) as a function of inverse temperature ($1/T$). Two different linear regimes are present, indicating that transport is via activation over two different barriers in thermally excited Wigner solids. The exact physical interpretation is unknown. The transition of the two regime takes place approximately at 300 mK.

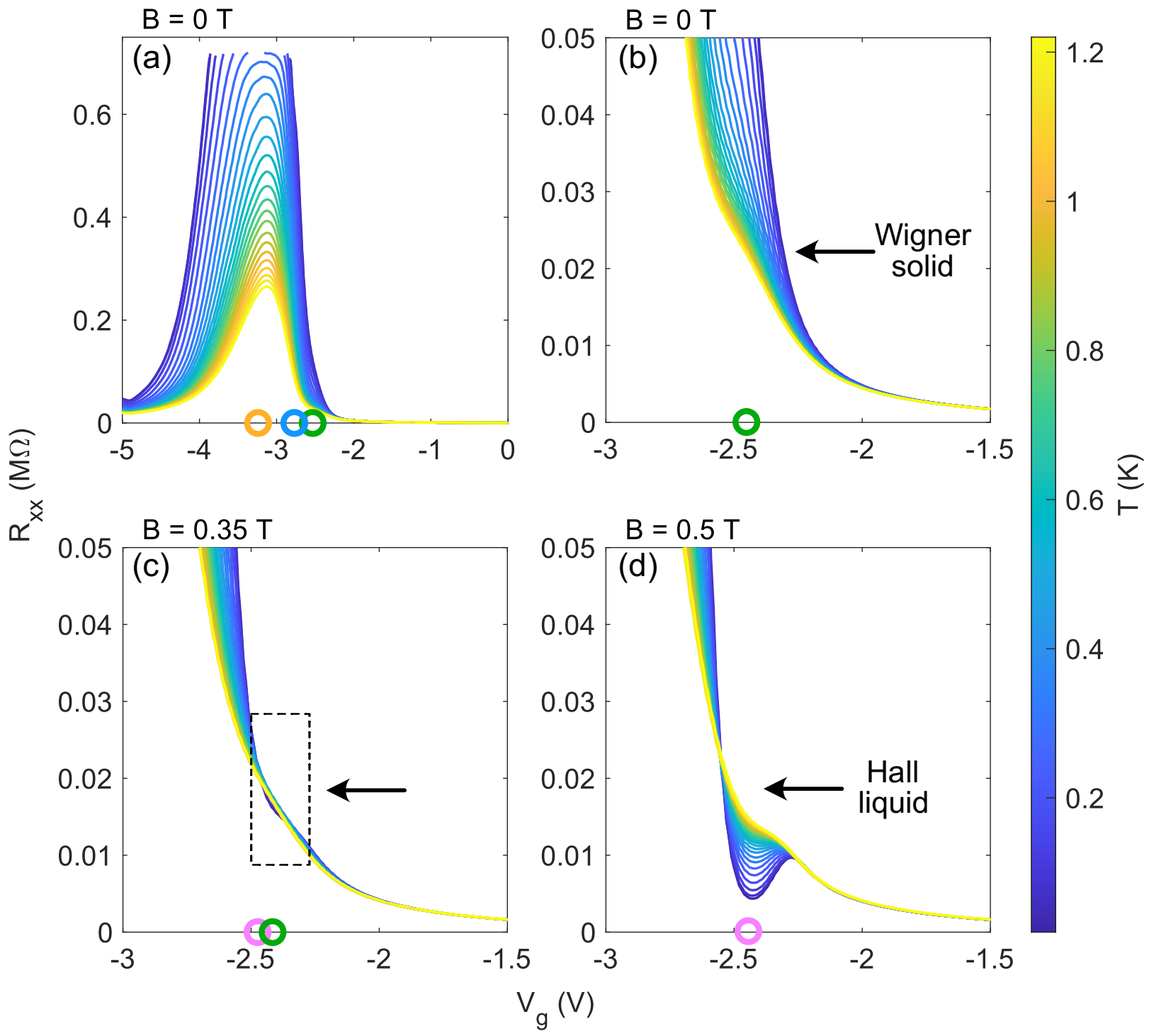

**Supplementary Figure 3:** Low field traces of the AC longitudinal resistivity ($R_{xx}$) as a function of the gate voltage at different temperatures. (a) Wider $V_g$ range resistivity data across the charge neutrality gap at $B = 0$. The circles indicate the approximate $V_g$ values at the centers of the charge neutrality gap (blue) and the electron (green) and hole (orange) Wigner solids, respectively. Zoom-in into the vicinity of electron Wigner solid at (b) $B = 0$ T (c) $B = 0.35$ T, and (d) $B = 0.5$ T. The emerging quantum Hall liquid corresponds to $\nu = 1$ and is labeled by pink circles. The region enclosed by a dashed box is shown in detail in Fig. 4 in the main text.

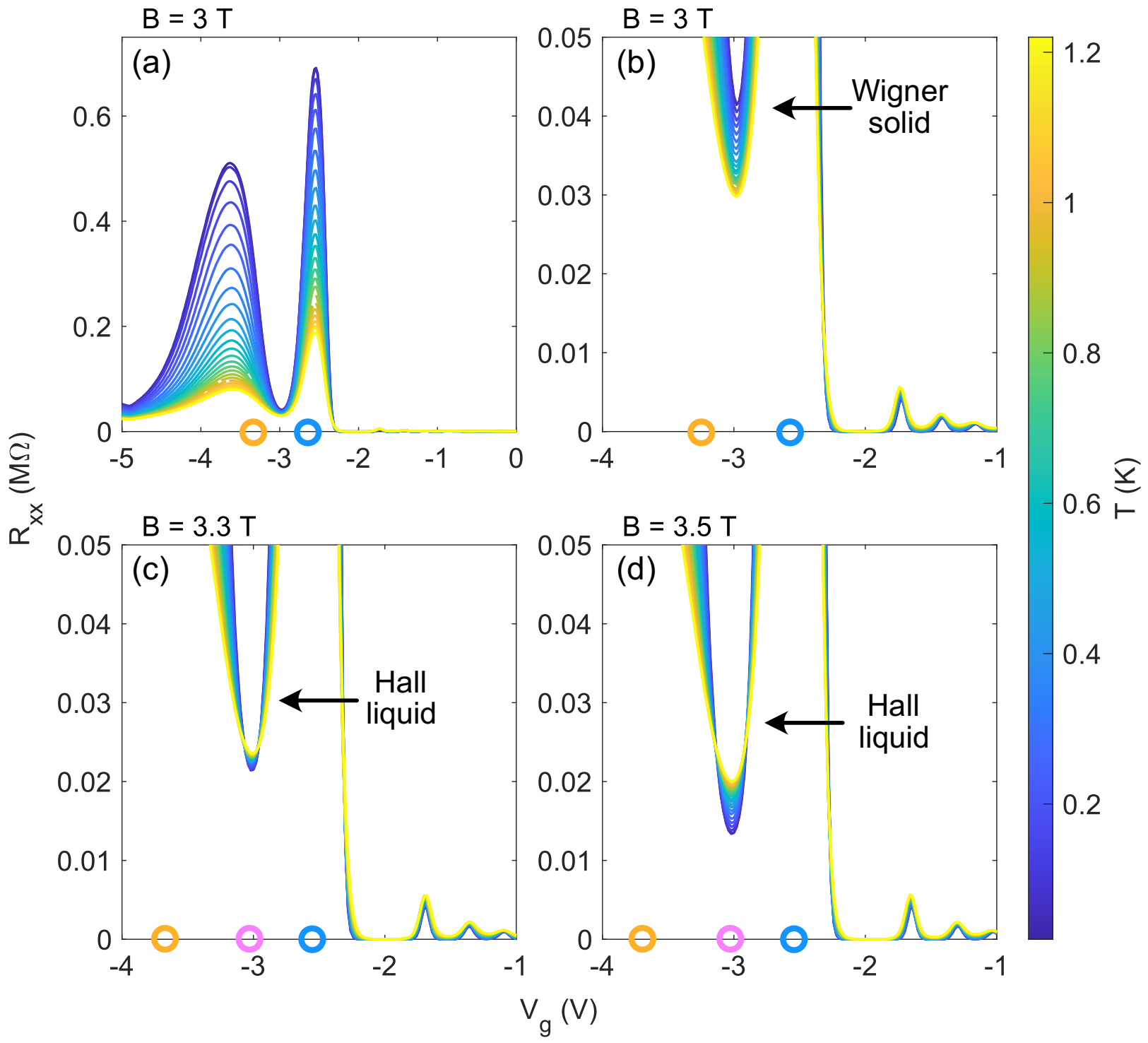

**Supplementary Figure 4:** High field traces of the AC longitudinal resistivity ($R_{xx}$) as a function of the gate voltage at different temperatures. (a) Wider $V_{\rm g}$ range resistivity data across the charge neutrality gap at $B$ = 3. The circles indicate the approximate $V_{\rm g}$ values at the centers of the charge neutrality gap (blue) and the hole (orange) Wigner solid, respectively. Zoom-in (b) $B$ = 3 T (c) $B$ = 3.3 T, and (d) $B$ = 3.5 T. The emerging quantum Hall liquid corresponds to $\nu = -1$ and is labeled by the pink circles.

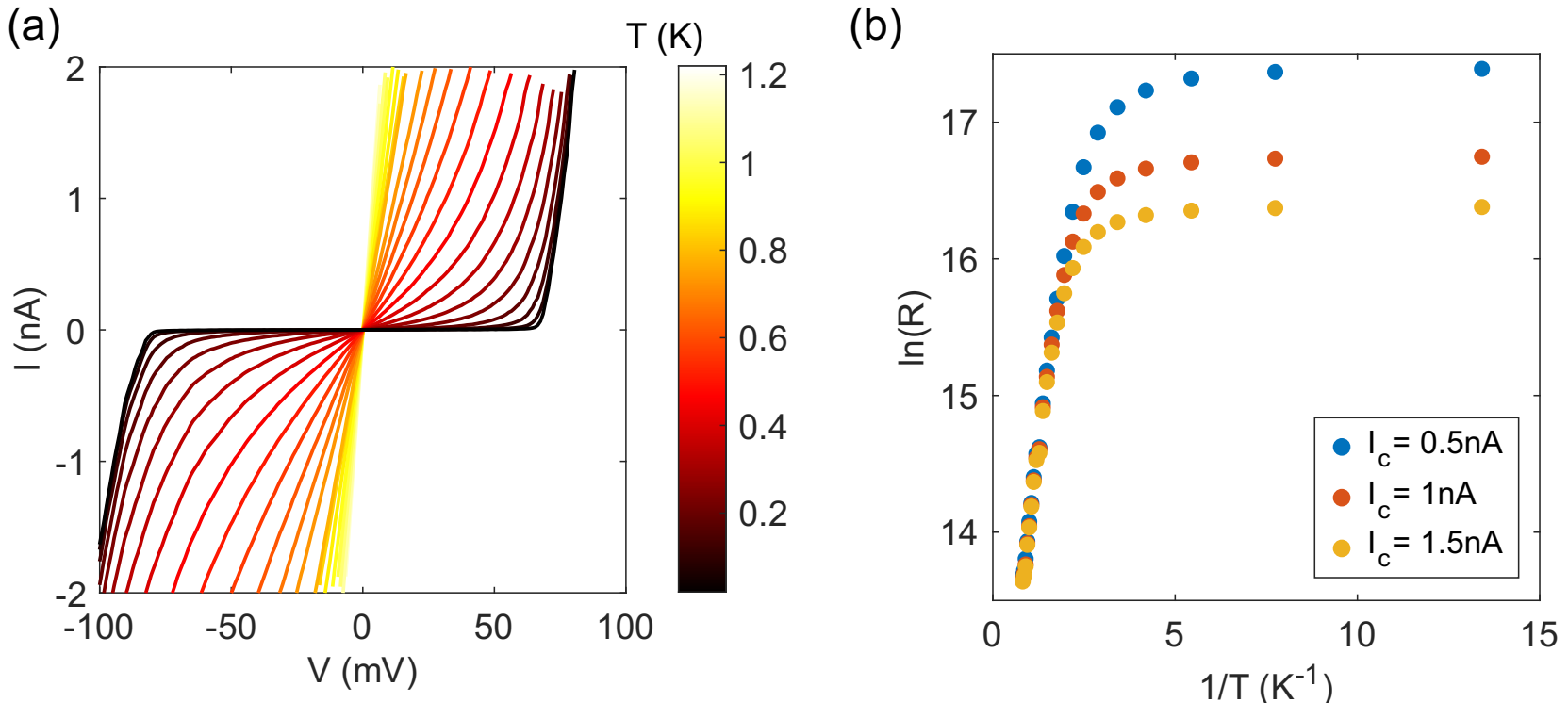

**Supplementary Figure 5:** (a) DC *I*-*V* traces of the hole Wigner solid at different temperatures *T*. The gate voltage is set to $V_{\rm g}$ = -3.6 V. (b) Logarithmic total resistance ($\ln R$) as a function of inverse temperature ($1/T$). Resistances $R$ were calculated for three different DC currents $I_{\rm c}$ (see labels).

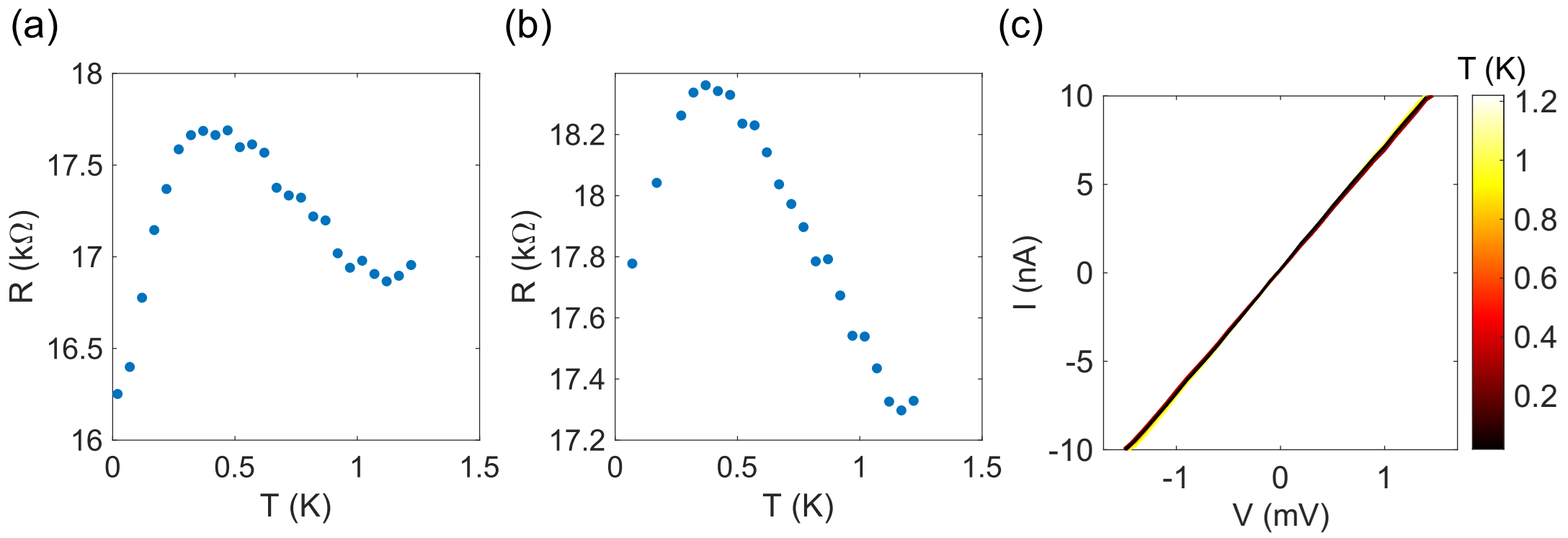

**Supplementary Figure 6:** Comparison of (a) AC resistivity and (b) DC resistivity. (c) DC I-V traces near Hall liquid-insulator transition ($B = 0.35$ T, $V_g = -2.36$ V). The difference between AC and DC resistivity is less than 1 kΩ. The metal-insulator transition around 400 mK can be detected in both measurements. The linear *I-V* allows for accurate AC measurements in this regime.